\documentclass{Interspeech2024}
\usepackage{subfigure}

\interspeechcameraready

\title{Global-Local Convolution with Spiking Neural Networks for Energy-efficient Keyword Spotting}

\name[affiliation={1}]{Shuai}{Wang}
\name[affiliation={1}]{Dehao}{Zhang}
\name[affiliation={1}]{Kexin}{Shi}
\name[affiliation={1}]{Yuchen}{Wang}
\name[affiliation={1}]{Wenjie}{Wei}
\name[affiliation={2}]{Jibin}{Wu}
\name[affiliation={1,*}]{Malu}{Zhang}

\address{
  $^1$University of Electronic Science and Technology of China\\
  $^2$The Hong Kong Polytechnic University }
\email{wangshuai718@std.uestc.edu.cn, jibin.wu@polyu.edu.hk, maluzhang@uestc.edu.cn}


\begin{document}

\maketitle

\begin{abstract}
Thanks to Deep Neural Networks (DNNs), the accuracy of Keyword Spotting (KWS) has made substantial progress. However, as KWS systems are usually implemented on edge devices, energy efficiency becomes a critical requirement besides performance. Here, we take advantage of spiking neural networks' energy efficiency and propose an end-to-end lightweight KWS model. The model consists of two innovative modules: 1) Global-Local Spiking Convolution (GLSC) module and 2) Bottleneck-PLIF module. Compared to the hand-crafted feature extraction methods, the GLSC module achieves speech feature extraction that is sparser, more energy-efficient, and yields better performance. The Bottleneck-PLIF module further processes the signals from GLSC with the aim to achieve higher accuracy with fewer parameters. Extensive experiments are conducted on the Google Speech Commands Dataset (V1 and V2). The results show our method achieves competitive performance among SNN-based KWS models with fewer parameters.
\end{abstract}

\renewcommand{\thefootnote}{\fnsymbol{footnote}}
\footnotetext{*Corresponding author.}

\section{Introduction}

Keyword Spotting (KWS) systems recognize predefined commands, which are always deployed on edge devices as an interface for human-machine interaction. Current mainstream KWS models implemented with Artificial Neural Networks (ANNs)~\cite{yang22n_interspeech} achieve outstanding accuracy. However, limited by endurance, CPU resources, and portability, deploying and running ANNs on edge devices for extended periods can be difficult.
Therefore, designing a high-accuracy, lightweight, and energy-efficient KWS model for edge devices is a hot topic that awaits a solution.

As the third-generation neural networks, Spiking Neural Networks(SNNs)~\cite{zhu2024tcja,zhang2021rectified} have gained widespread attention due to their asynchronous event-driven architecture~\cite{akolkar2015can} and ultra-low energy consumption.
The spiking event-driven mechanism~\cite{wei2024event} of SNNs computes only when necessary, resulting in sparse information transmission and significantly reduced energy consumption. This is suitable for resource-constrained edge devices. Besides, it has been proved that accumulate
(AC) operation is compact and energy-efficient compared with Multiply-and-Accumulate (MAC) operation~\cite{karmakar2017impact}. Therefore, when deployed on hardware, SNNs using AC operation expend much less energy in comparison with MAC-dependent DNNs~\cite{maass1997networks}.

The advantages of SNNs have motivated many researchers to apply them to KWS tasks~\cite {pan2020efficient,wu2018spiking}. 
However, many attempts based on deep SNNs still use FFT~\cite{kim202223muw}, MFCC~\cite{tiwari2010mfcc} to pre-process raw speech (wav) which needs massive computing resources. 
This goes against to our original intention of using SNNs to implement energy-efficient KWS models. To avoid this problem, some researchers have attempted to directly utilize raw speech signals with low-resource-consuming convolutional operations~\cite{phiphitphatphaisit2021deep}. For instance,  Philipp et al.~\cite{weidel2021wavesense} proposed the end-to-end streaming model by dilated convolution with $stride=1$. However, this approach fails to compress the length of long speech sequences, leading to evident feature redundancy. What's more, Yang et al.~\cite{yang2022deep} proposed an end-to-end deep residual SNN model, and it demonstrated a notably high level of recognition accuracy. However, their approach employs the integrate-and-fire (IF) neuron model, which lacks a membrane potential decay mechanism. This will result in frequent spikes firing, leading to higher computational energy consumption in SNNs\cite{fang2021incorporating}.

\begin{figure*}[htbp]
 \centering
 \subfigure[]{
        \label{Fig.sub.2}
 	\includegraphics[scale=0.53]{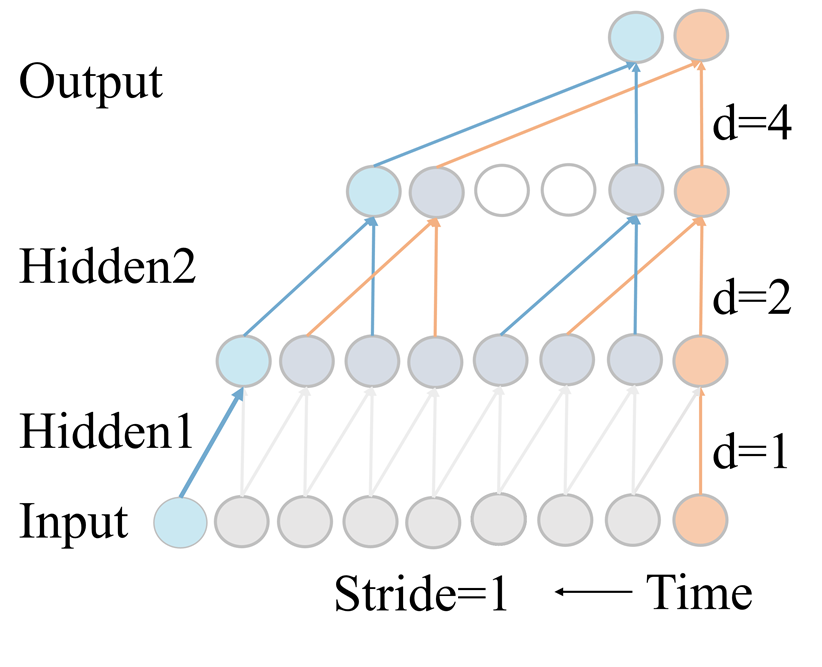}
 }
 \subfigure[]{
        \label{Fig.sub.3}
 	\includegraphics[scale=0.53]{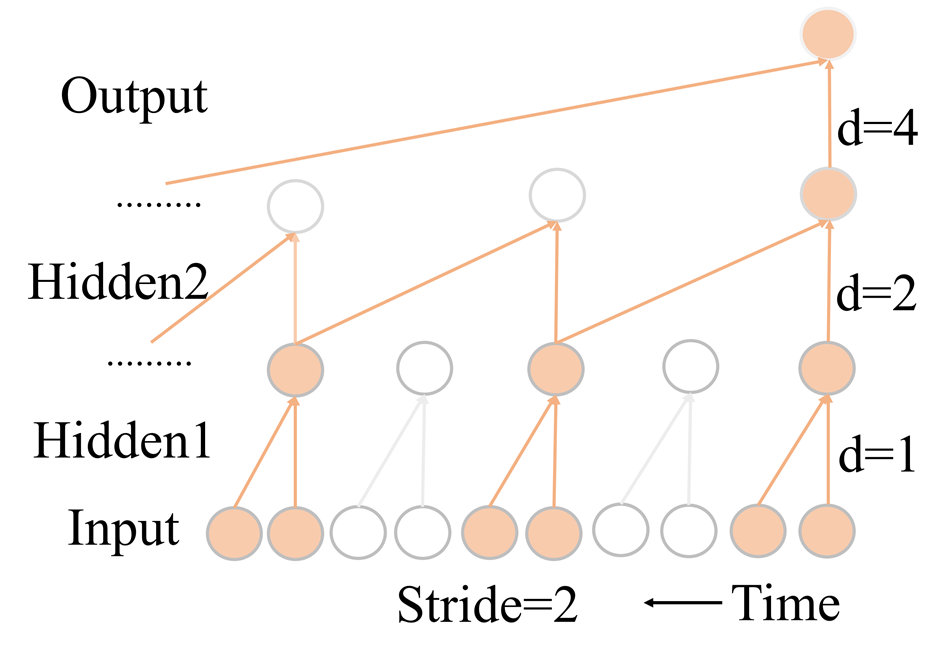}
  }
  \subfigure[]{
        \label{Fig.sub.LS}
 	\includegraphics[scale=0.53]{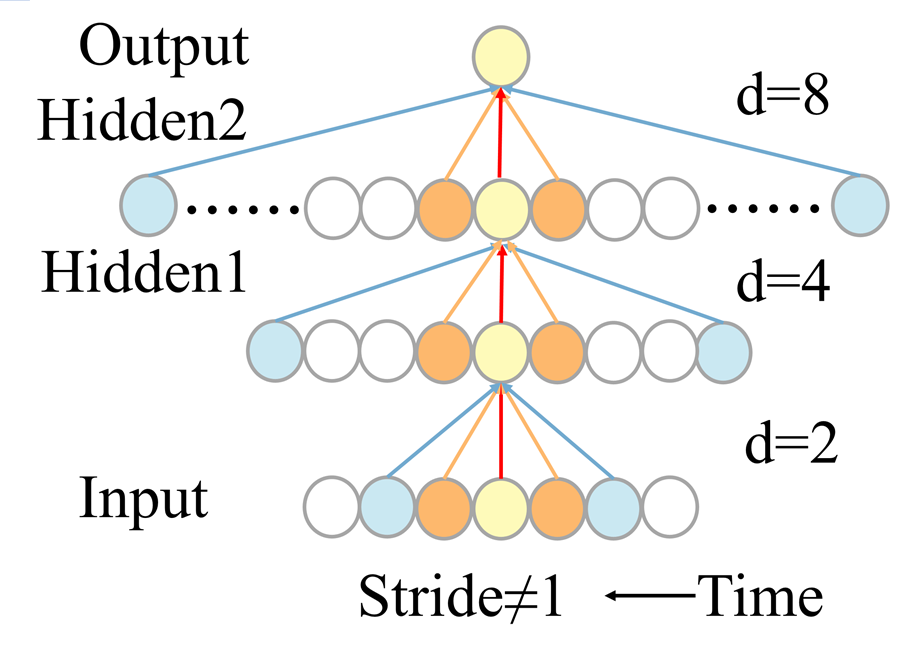}
  }
  \subfigure[]{
        \label{Fig.sub.sin}
 	\includegraphics[scale=0.24]{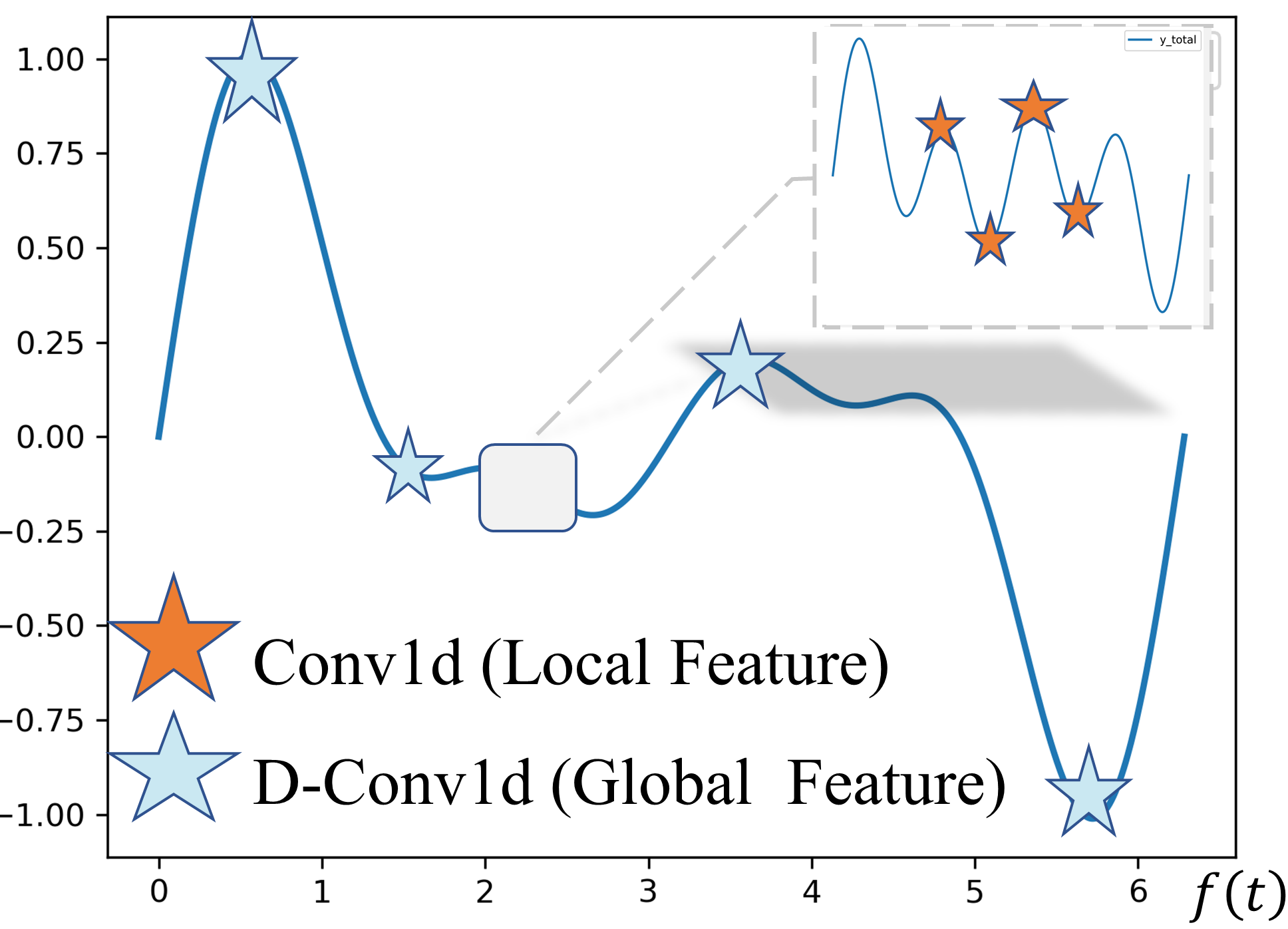}
  }
 \caption{A comparative analysis between single convolution and global-local convolution. (a) Dilated Conv1d with $ stride=1$. The hidden layers and output features are highly redundant, as evidenced by the gray blocks representing the overlapping features. (b) Dilated Conv1d with $ stride\neq1$. The receptive field exponentially increases with the dilation factor $d$, leading to a loss of local information as white blocks. (c) and (d) the Global-Local convolution method. 
 it can achieve a good balance between global and local features in long speech sequences, and maintain a consistent focus on local features when the $ stride\neq1$.
 }
 \label{fig:distance}
\end{figure*}

In this paper, we constructed an end-to-end SNN-based KWS model to address the issues above. We tested the accuracy of our model on the Google Speech Commands datasets~\cite{warden2018speech} (V1 and V2) and compared its model size with the related works based on SNNs. Encouragingly, compared with similar SNN-based models,
our model achieves competitive performance on smaller model sizes.
Finally, energy efficiency calculations prove that our model consumes 10$\times$ less energy than ANNs with the same structure. Hence, our SNN-KWS model aligns perfectly with the requirements of edge devices for high accuracy, lightweight structure, and low energy consumption. The major contributions of this paper can be summarized as follows:
\begin{itemize}
\item Global-Local Spiking Convolution (GLSC) module: We design the GLSC module to achieve better and more energy-efficient spiking convolution. It can compress the length of long speech sequences layer by layer while considering both global and local features.
\item Bottleneck-PLIF module:
To achieve a more lightweight and efficient SNN architecture, we combine the Bottleneck structure in ResNet~\cite{he2016deep} with more efficient Parametric Leaky Integrate-and-Fire (PLIF)~\cite{fang2021incorporating} neurons to create a more lightweight classifier.
\item By integrating the proposed GLSC and Bottleneck-PLF modules, we construct a novel end-to-end SNN-KWS model. Our SNN-KWS model achieves competitive performance in both accuracy and parameter efficiency within the domain of SNN-based models.

\end{itemize} 

\section{Preliminaries}
In this section, we will give an overview of two essential components in our model: end-to-end speech feature extraction and spiking neural networks. Additionally, we will analyze the challenges associated with these components.
\subsection{End-to-end Convolution for Speech Features}

To alleviate the energy consumption caused by conventional speech feature extraction methods, such as  FFT~\cite{kim202223muw}, MFCC~\cite{tiwari2010mfcc}, the most popular approach is to use direct convolution methods for end-to-end feature extraction. They can be summarized as:
\begin{equation}
f\left(t\right)\ast g\left(t\right)=\int_0^tf\left(u\right)g\left(t-u\right)du
\end{equation} 
$ g\left(t\right)$ is regarded as different convolution kernels and $ f\left(t\right)$ is denoted as original speech wave sequences. While the convolution methods have been successfully applied in speech feature extraction, certain challenges still require further resolution.

For example, the parameters in Conv1d~\cite{johnson2017deep} increase with the expansion of the receptive field, resulting in the redundancy of parameters in $g\left(t\right)$.
The dilated Conv1d (D-Conv1d)~\cite{DBLP} was proposed to address this problem. However, as illustrated in Figs \ref{Fig.sub.2} and \ref{Fig.sub.3}, dilated convolution~\cite{weidel2021wavesense} with $stride=1$ suffers from redundancy of features and $ stride\neq1$ may lead to loss of local features. Therefore, there is no convolution method currently that can simultaneously address both the redundancy of parameters and the loss of local features. 

\subsection{Spiking Neural Networks}
SNNs encode information through binary spikes over time and work in an event-driven manner, which has a great advantage in energy consumption. As a basic unit of SNNs, various spiking neurons are proposed to emulate the mechanism of biological neurons. Among them,  the Leaky Integrate-and-Fire (LIF) \cite{wu2019direct} model is widely used due to its simplicity. The dynamics of a LIF neuron can be expressed as follows:
\begin{equation}
U_i^{t+1,n}=\tau U_i^{t,n}+\sum_{j=1}^{l\left(n-1\right)}w_{ij}^nO_j^{t+1,n-1}\label{6}
\end{equation}
where $\tau$ is the constant leaky factor, $U_i^{t+1,n}$ is the membrane potential of neuron $i$ in the $n$th layer at the time step $t+1$, and $\sum_{j=1}^{l\left(n-1\right)}w_{ij}^nO_j^{t+1,n-1}$ denotes
the pre-synaptic inputs for neuron $i$.
When the membrane potential $U_i^{t+1,n}$ exceeds the firing threshold $V_{th}$, the neuron 
$i$ fires a spike $O_i^{t+1,n}$ and $U_i^{t+1,n}$ reset to 0. The firing function and hard reset mechanism can be described by Eq.~\ref{7} and Eq.~\ref{8}, respectively.
\begin{equation}
O_i^{t+1,n}=H\left(U_i^{t+1,n}-V_{th}\right)\label{7}
\end{equation}
\begin{equation}
U_i^{t+1,n}=U_i^{t+1,n}(1-O_i^{t+1,n}) \label{8}
\end{equation}
where $H$ denotes the Heaviside step function. 

Many studies have effectively leveraged the energy efficiency of spiking neurons to develop energy-efficient SNN-KWS models~\cite{wang2023spatial,zhang2023long}. However, these methods have not taken into account the lightweight structural requirements of edge devices. Therefore, we aim to design a more lightweight and energy-efficient SNN-KWS model by utilizing more advanced structures and spiking neuron models.
\section{Method}
\begin{figure}[htpb] 
    \centering 
    \includegraphics[scale=0.3]{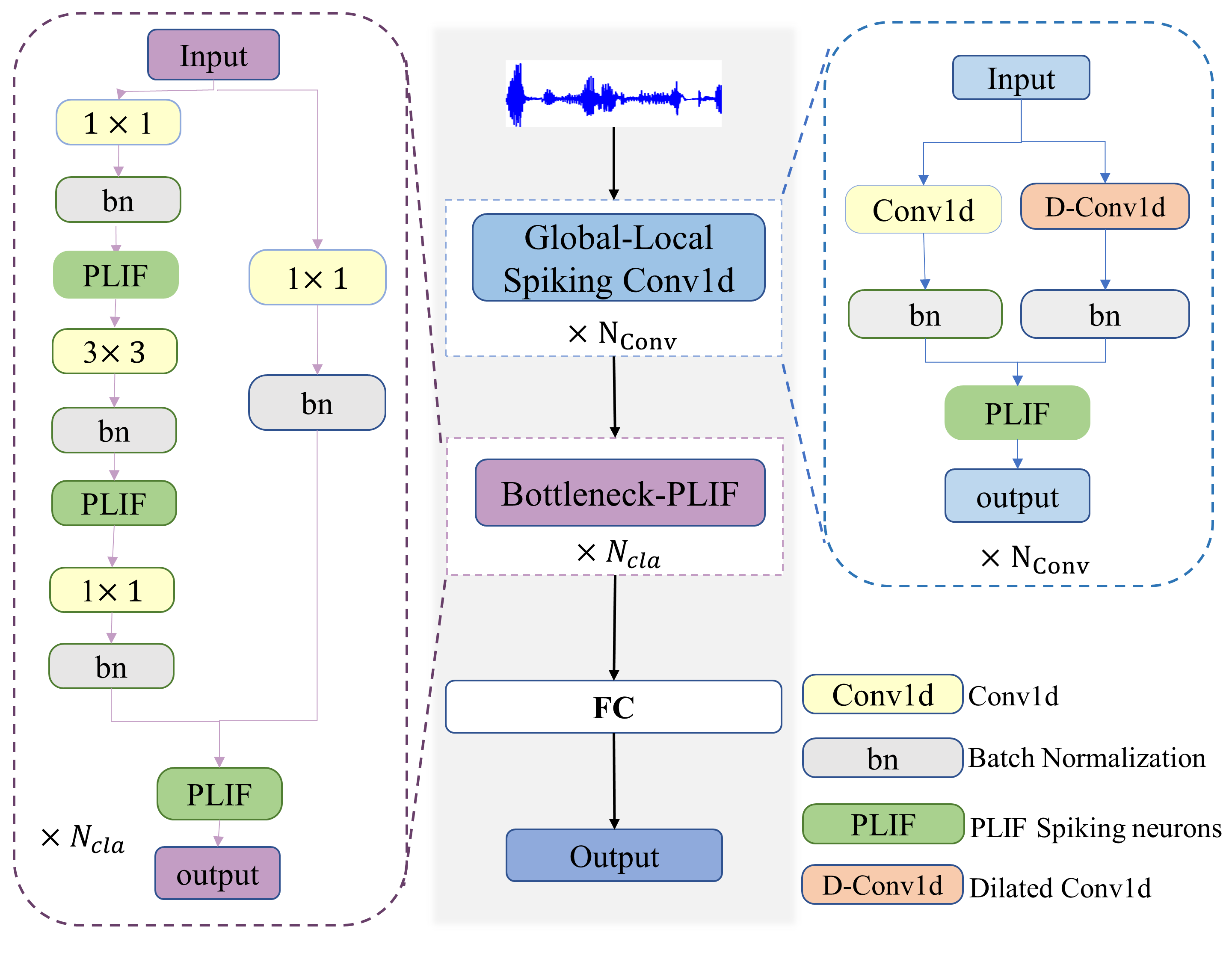}
    \caption{Our SNN-KWS model structure. It consists of $ N_{Conv}=4 $ GLSC blocks (right part) for better feature extraction, and $ N_{Cla}=2 $ Bottleneck-PLIF blocks (left part) for effective classification. }
    \label{picture1}
\end{figure}

In this section, we propose an end-to-end SNN-KWS model that effectively addresses the limitations mentioned in Section 2. The overall structure of the model is illustrated in Fig.\ref{picture1}, which mainly comprises two innovative modules: 1) the GLSC module and 2) the Bottleneck-PLIF module. 

\subsection{Global-Local Spiking Conv1d Module }
To achieve better and energy-efficient speech feature extraction, we propose a Global-Local Spiking Conv1d module for end-to-end feature extraction. GLSC mainly consists of three components, Conv1d, D-Conv1d, and spiking neurons. The flowchart of the GLSC module is illustrated in the right portion of Fig. \ref{picture1}, and it can be  mathematically expressed as:
\begin{equation}
outputs=H\left(bn\left(g_1\left(t\right)\ast f\left.\left(t\right.\right)\right)+ bn\left(g_2\left(t\right)\ast f\left(t\right)\right)\right)
\end{equation}
where $ g_1 $, $ g_2 $ are the convolution kernels of Conv1d and D-Conv1d, respectively. $bn$ is Batch Normalization and $ H$ is the firing function of spiking neurons as Eq.\ref{7}. 
In the following, we will analyze how the proposed GLSC can achieve enhanced feature extraction and energy efficiency.

\begin{figure}[htpb] 
    \centering 
    \label{Fig.sub.11}
    \includegraphics[scale=0.31]{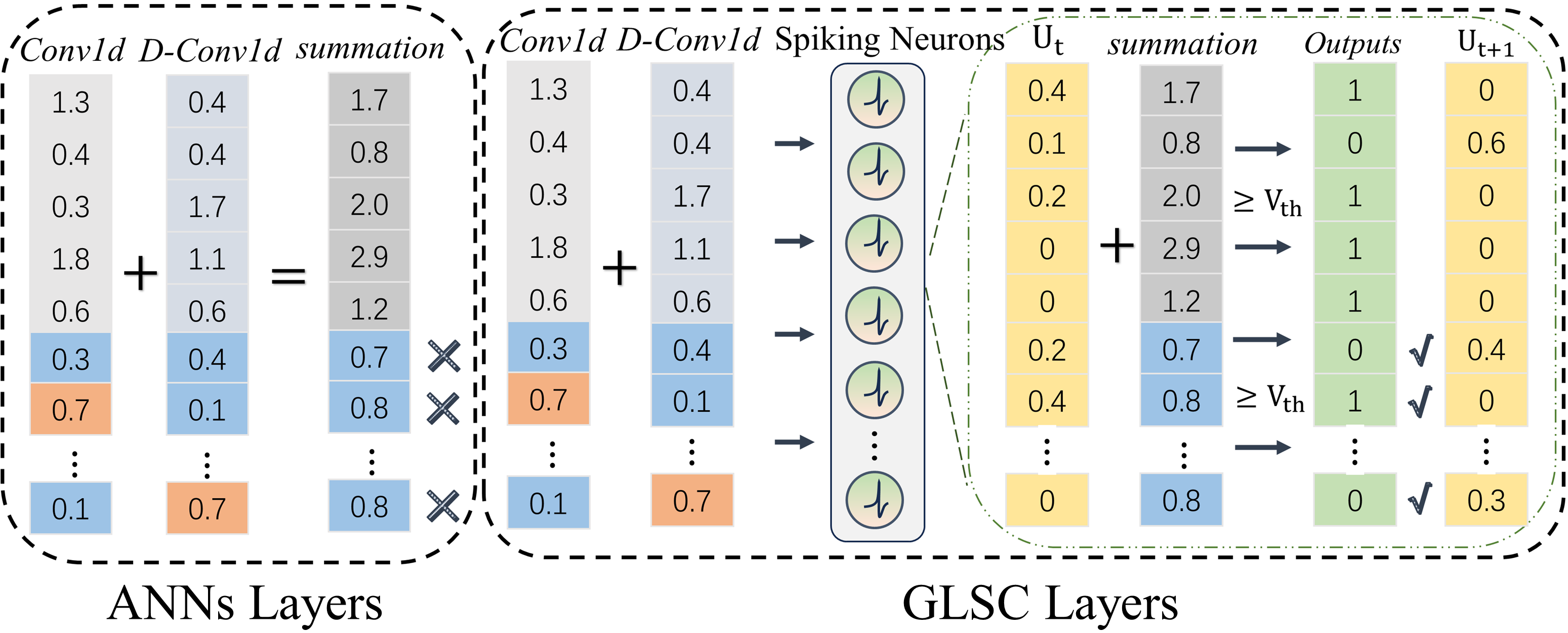}
    \centering
    \caption{The Global-local convolution feature extraction in ANNs and GLSC layers. $ U_{t+1}$ represents the membrane potential contribution of spiking neurons after decaying from $ U_{t}$. 
      }
    \label{picture4}
\end{figure}

In contrast to a single Conv1d or D-Conv1d method, the proposed GLSC module can benefit from both worlds. As demonstrated in Fig.~\ref{Fig.sub.sin}, the GLSC module can effectively balance local and global features in long speech sequences. Although the idea of global-local feature extraction exists in some ANN-based models such as Branchformer~\cite{peng2022branchformer}, their success relies on utilizing complex attention mechanisms rather than combining global and local convolutions directly. In ANNs, merging two convolutions directly leads to feature disappearance, where salient details like the orange block become indistinct upon addition as shown in Fig.\ref{picture4}.

Here, we innovatively employ spiking neurons to solve these problems and achieve simpler and sparser global-local feature extraction. As shown in the right part of Fig.\ref{picture4}, the output of spiking neurons at $t$ depends not only on the summation but also considers the residual membrane potential from $t-1$, significantly mitigating feature disappearance caused by the summation. We will validate this aspect through ablation studies.
Moreover, only when the $summation + U_{t}$ is greater than the $V_{th}$, can spiking neurons pass key information for outputs as the green blocks. This approach effectively prevents the accumulation of irrelevant features across layers, while ensuring that the feature vectors become sparser.
So the GLSC module achieves a sparser end-to-end feature extraction.

\subsection{ Bottleneck-PLIF Module}
\begin{figure}[htpb] 
    \centering 
     \includegraphics[scale=0.24]{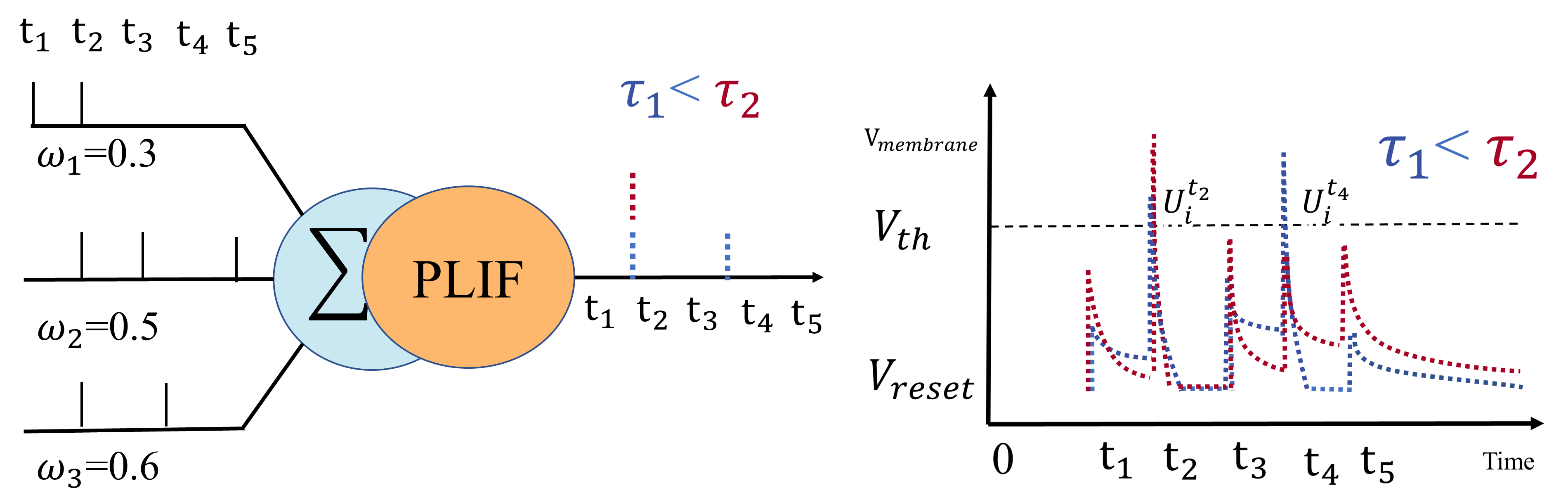}
    \caption{with the same inputs, these neurons with different $\tau$ result in varied leaky rates for neurons' membrane potential(right part), thereby leading to diverse output results(left part). }
    \label{figplif}
\end{figure}

To address the limitations of the existing SNN-based KWS  models, we take advantage of the effective PLIF spiking neuron and lightweight bottleneck structure to construct the Bottleneck-PLIF module.
\begin{equation}
U_i^{t+1,n}=U_i^{t,n}-k\left(a\right)\left(\ U_i^{t,n}-\ \sum_{j=1}^{l\left(n-1\right)}w_{ij}^nO_j^{t+1,n-1}\right)\label{12}
\end{equation} 
Eq.~\ref{12} depicts the membrane potential dynamics of a PLIF neuron.
Compared to traditional LIF neurons, the PLIF neuron exhibits two notable enhancements.
First, learnable $k\left(a\right)$ replaces the constant decay hyperparameters $\tau$ in Eq.\ref{6}, which can be optimized during training. Second, PLIF applies learnable $ k\left(a\right)$ to the input. As depicted in Fig.\ref{figplif}, neurons exhibit a greater diversity of outputs when subjected to different $\tau$ under the same input conditions.
 
Meanwhile, inspired by the Bottleneck block in ResNet \cite{he2016deep}, it can efficiently integrate feature information with fewer parameters and reduce feature dimensionality by fusing channels. We incorporate PLIF into the Bottleneck structure to achieve a more efficient spiking classifier as shown in the left part of Fig. \ref{picture1}, and its mathematical expression is as follows:
\begin{equation}
Outputs=H\left[f_1(H(f_3(H(f_1(Input))+f_1(Input)\right]\label{13}
\end{equation}
$ f_n$  represents $ n\times n$ convolution and Batch Normalization, and $ H$ represents the firing function of PLIF. In Eq.\ref{13}, $ f_1$ are used for fusing channels without compromising the input features, which can reduce feature dimensionality without compromising the original structure.$ f_3$ are used to further computer the previous spikes features, which can further process the signals from the GLSC module with a more lightweight model size. 

\section{Experiments}
\subsection{Dataset}
The Google Speech Commands (GSC)~\cite{warden2018speech} dataset includes 30 short commands for Version 1 (V1) and 35 for Version 2 (V2), recorded by 1,881 and 2,618 speakers, respectively. 
To make a fair comparison, our experiments are conducted on the 12-class and 35-class classification tasks as previous SNN models~\cite{yang2022deep, orchard2021efficient}. While 12-class classification recognizes 12 classes, that include 10 commands: “yes”, “no”, “up”, “down”, “left”, “right”, “on”, “off”, “stop” “go”, and two additional classes: silence, and an unknown class. The unknown class covers the remaining 20 (25) speech commands in the set of 30 (35). The silence class accounting for about 10 $\%$ of the total dataset is generated by splicing the noise files in the dataset. Finally, GSC-V1 is split into 56588 training, 7743 validation, and 7835 test utterances, and GSC-V2 is divided into 92843 training, 11003 validation, and 12005 test utterances. We use the STBP\cite{wu2018spatio} method to train the entire model directly.

 \subsection{ Accuracy and Model Size}
To validate the accuracy and model size of our proposed model, we conduct a comprehensive comparative analysis with previous studies~\cite{pellegrini2021low,yang2022deep,weidel2021wavesense,orchard2021efficient,wang2023spatial,zhang2023long, yin2021accurate,stewart2023speech2spikes}. The experimental results are shown in Table \ref{tab:word_styles}. 
Although our accuracy is slightly lower than ST-Attention-SNN and SRNN+ALIF, our model size is significantly smaller. In conclusion, our KWS-SNN achieves competitive performance in both 12-class and 35-class tasks with a substantially reduced model size. This indicates that our model can be easier to deploy on edge devices.
  \begin{table}[t]
  \caption{A summary of KWS models' accuracy and model size.}
  \label{tab:word_styles}
  \centering
  \begin{tabular}{c c c}
    \toprule
    \textbf{Model}     &\textbf{Model Size(K)}   &  \textbf{Acc(\%)}\\
    \midrule
    \multicolumn{3}{c}{Google Speech Commands Dataset Version 1 (12)}\\[1ex]
    NLIF full SNN\cite{pellegrini2021low}         & 220  & 87.9 \\
    E2E residual SNN\cite{yang2022deep}      &86.5      & 92.2 \\
    (Our) SNN-KWS          &\textbf{70.1}      &\textbf{93.0} \\
    \midrule
    \multicolumn{3}{c}{Google Speech Commands Dataset Version 2 (12)}\\[1ex]
    ST-Attention-SNN \cite{wang2023spatial}          &2170  &95.1\\
    SLAYER-RF-CNN \cite{orchard2021efficient}     &280      & 91.4  \\
    SpikGRU\cite{dampfhoffer2023leveraging}         &111      & 94.9 \\
    (Our) SNN-KWS              &\textbf{70.1}      &\textbf{94.4} \\
    \midrule
    \multicolumn{3}{c}{Google Speech Commands Dataset Version 2 (35)}\\[1ex]
    WaveSence \cite{weidel2021wavesense}       &N/A  &79.5\\
    LSTMs-SNN \cite{zhang2023long}          &N/A  &91.5\\
    SRNN+ALIF \cite{yin2021accurate}     &222.1      & 92.5  \\
    Speech2Spikes\cite{stewart2023speech2spikes}        &410      & 89.5 \\
    (Our) SNN-KWS            &\textbf{80.2}      &\textbf{92.9} \\
    
    \bottomrule
  \end{tabular}
\end{table}
\subsection{ Energy Efficiency}
\begin{figure}[htpb] 
    \centering 
    \includegraphics[scale=0.27]{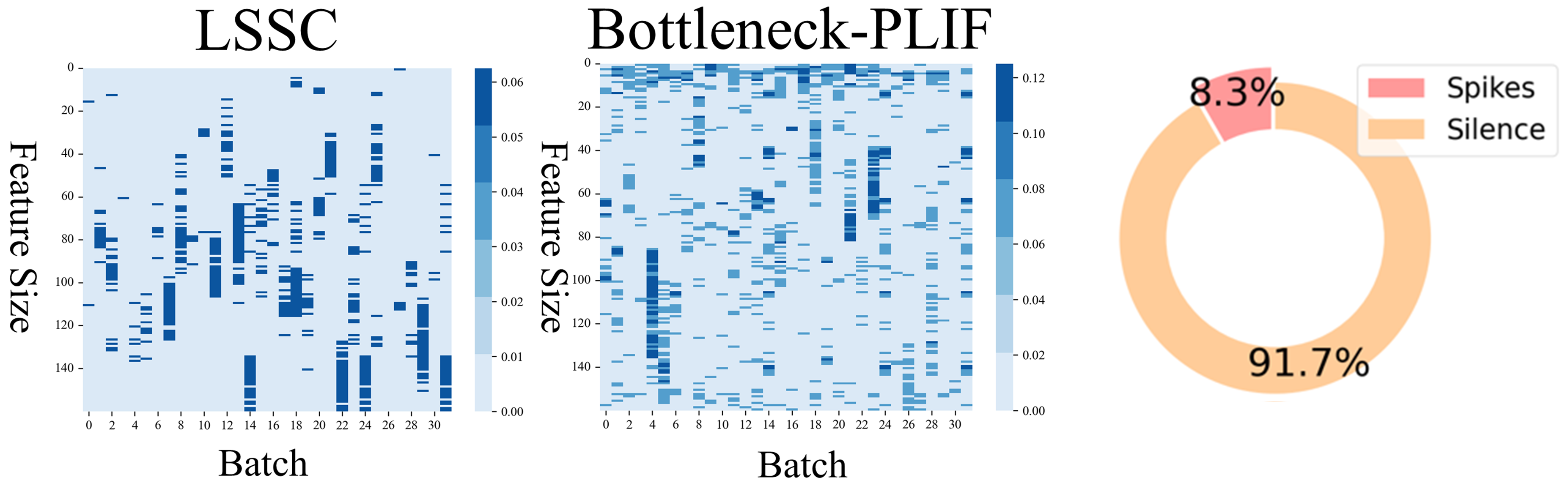}
    \caption{The average spike firing rate of our SNN-KWS model when $TimeSteps$ is 8 on the GSC-V1 dataset. The average spike firing rate of the entire network is approximately 8.3\%. }
    \label{picture2}
\end{figure}

In this part, we validate the energy efficiency advantage of our model over their ANNs counterparts. According to the standards established in the field of neuromorphic computing~\cite{sengupta2019going}, the energy consumption ratio between our model and an equivalent ANN model can be calculated as:
\begin{equation}
{Energy}_{rate}=\frac{AC}{MAC}\ \ast\ SpikingRate\ \ast\ TimeSteps\label{18}
\end{equation}
$\frac{AC}{MAC} $ is denoted as the energy consumption ratio between float-point additions(AC) in SNNs and float-point multiplications(MAC) in ANNs. Extensive research has substantiated that 
$\frac{AC}{MAC}=\frac{1}{7}$ ~\cite{horowitz20141}. $SpikingRate$ and $TimeSteps$ represent the average firing rate and simulation time window.
As illustrated in Fig.\ref{picture2}, the average spike firing rate of each module is $ 8.3\%$ and the $TimeSteps$ in our model is set to $ 8$. Therefore, according to Eq.\ref{18}, our SNN-KWS model achieves more than 10$\times$ energy saving over the ANNs counterpart.
\subsection{Ablation Study}
\begin{figure}[htpb] 
    \centering
  \subfigure[]{
        \label{Fig.sub.qqq}
 	\includegraphics[scale=0.42]{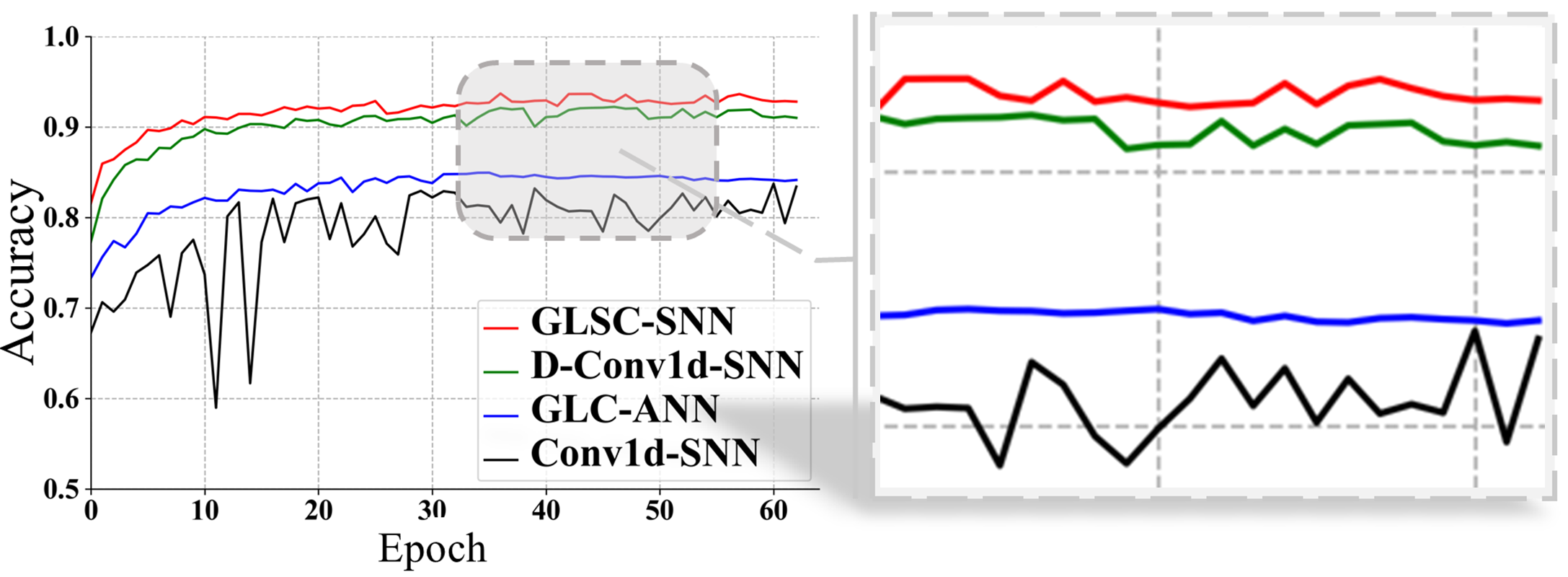}
   }  
 \subfigure[]{
        \label{Fig.sub.7}
 	\includegraphics[scale=0.49]{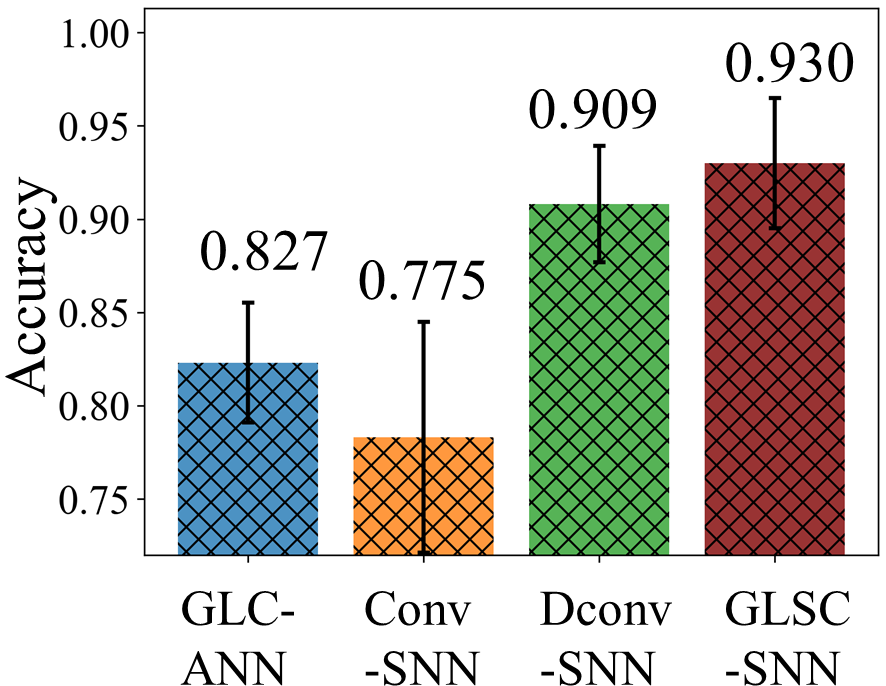}
   }
 \subfigure[]{
        \label{Fig.sub.8}
 	\includegraphics[scale=0.49]{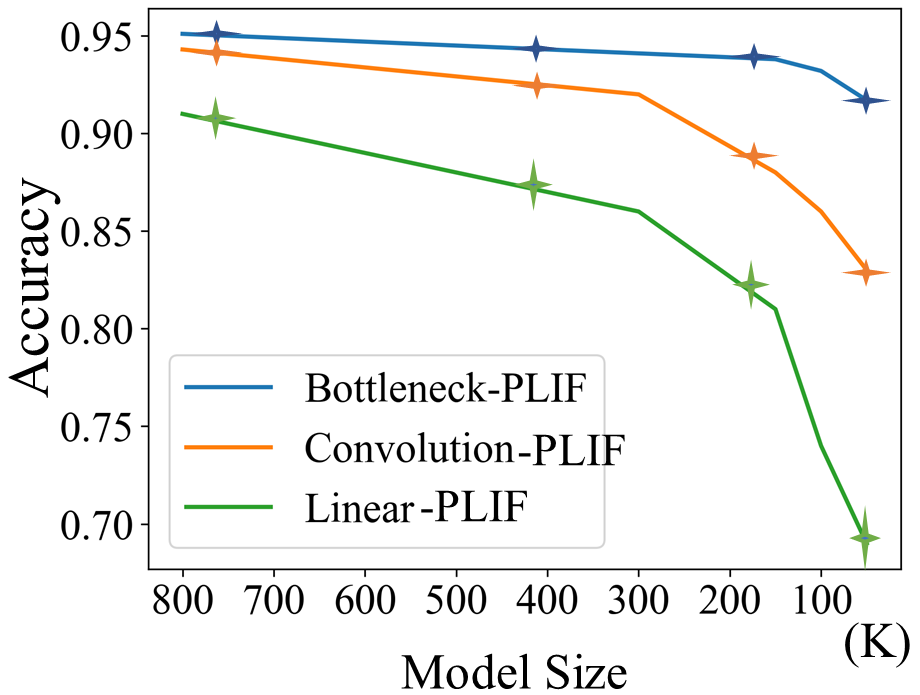}
 }
    \caption{Ablation Studies. (a,b)Validating the feature extraction capabilities of the GLSC module. (c) The performance advantage of the Bottleneck-PLIF module becomes more pronounced as the number of parameters decreases.}
    \label{picture6}
\end{figure}

In this part, we conduct ablation studies to validate the effectiveness of the GLSC and Bottleneck-PLIF modules, respectively. Firstly, we evaluate the GLSC by comparing it with single convolution methods on the same number of parameters. As illustrated in Figs.\ref{Fig.sub.qqq} and \ref{Fig.sub.7}, the GLSC module consistently surpasses other methods (black and green), exhibiting both better performance and convergence. It is noteworthy that the GLC-ANN(blue curve) represents the substitution of spiking neurons in the GLSC module with continuous activation functions of ANNs. By comparing the red and blue curves, it can be proven that spiking neurons play a key role in addressing the issue of feature disappearance.

Next, we verify that the Bottleneck module can allow us to achieve better performance while utilizing fewer parameters. As shown in Fig.\ref{Fig.sub.8},
the performance of all classifiers exhibits a decline as parameters decrease. However, the reduction in parameters has minimal impact on our Bottleneck-PLIF model, and our method can achieve an accuracy of 93\% even when the parameters are below 100K. 
\section{Conclusion}
In this work, we propose a novel SNN-KWS model with two innovative modules.
The GLSC module enhanced end-to-end convolution speech feature extraction. It avoids the high computation costs associated with traditional data pre-processing~\cite{kim202223muw,tiwari2010mfcc}, while simultaneously considering both global and local speech features. 
The Bottleleck-PLIF module further calculates the spike features from the GLSC module, with the aim of achieving higher classification accuracy using fewer parameters. 
By conducting experiments on the GSC~\cite{warden2018speech} dataset, our model achieves competitive performance in both accuracy and parameter efficiency among similar SNN-based models and achieves more than 10× energy saving over the ANNs.
Therefore, our SNN-KWS model proficiently satisfies the requirements of edge devices in terms of exceptional accuracy, lightweight design, and energy efficiency. In the future, we will implement
it realistically on a neuromorphic chip.

\section{Acknowledgements}
This work was supported by the National Science Foundation of China under Grant 62106038, and in part by the Sichuan Science and Technology Program under Grant 2023YFG0259.
\bibliographystyle{IEEEtran}
\bibliography{mybib}

\end{document}